\documentclass[reprint,aps]{revtex4-1}

\usepackage[dvips]{graphicx}
\usepackage{amsmath,amssymb,yhmath,makeidx,mathdots,setspace,framed}
\usepackage{dcolumn}
\usepackage{bm}

\begin{document}

\preprint{APS/123-QED}

\title{Thermodynamics of the exactly solvable spin-electron tetrahedral chain}

\author{L. G\'{a}lisov\'{a}}
\email{galisova.lucia@gmail.com}
 \affiliation{Department of Applied Mathematics and Informatics,
             Faculty of Mechanical Engineering, Technical University in Ko\v{s}ice,
             Letn\'{a} 9, 042 00 Ko\v{s}ice, Slovak Republic}


\begin{abstract}
An exactly solvable spin-electron tetrahedral chain, where the Ising spins localized at nodal lattice sites regularly alternate with three equivalent lattice sites available for one mobile electron is considered. The system with ferromagnetic interaction between the Ising spins and electrons exhibits an enhanced magnetocaloric effect (MCE) in the limit $H/|J|\to 0$ when the entropy is very small, whereas the system with antiferromagnetic interaction between the Ising spins and electrons exhibits an enhanced MCE around the field $H/J=1$ when the entropy is sufficiently close to the value $S/2N = \ln[(1+\sqrt{5})/2]$. We study the thermodynamics of the system in these field regions.
\begin{description}
\item[PACS] 05.70.-a; 75.10.Pq
\item[DOI] 10.12693/APhysPolA.volume.page
\end{description}
\end{abstract}

\maketitle

\section{Introduction}
Exactly solvable 1D models provide an excellent playground for theoretical studies of collective phenomena. One such model is a coupled tetrahedral chain, which has been exactly solved for the spinless fermion model~\cite{1} and the mixed Ising-Heisenberg model~\cite{2}. In this paper, we consider another exactly solvable model with tetrahedral-chain structure, where the Ising spins localized at nodal lattice sites alternate with three equivalent lattice sites available for mobile electrons. Note that the particular case of this hybrid model with one mobile electron delocalized over each triangular plaquette has been recently solved to study the ground state, the magnetization process and the magnetocaloric effect of the system~\cite{3}. The aim of the present work is to shed light on the thermodynamics of the model in those field regions, where an enhanced MCE has been found.

\section{Spin-electron tetrahedral chain}
\label{sec:model}

\begin{figure}[ht]
\begin{center}
\vspace{-0.75cm}
\includegraphics[angle = 0, width = 0.85\columnwidth]{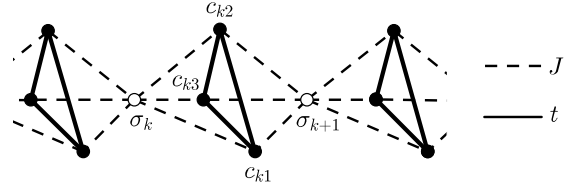}
\vspace{-0.1cm}
\caption{\small A part of the spin-electron system on tetrahedral chain. The white (black) circles indicate the lattice sites available for localized Ising spins (mobile electrons).}
\label{fig1}
\end{center}
\vspace{-0.25cm}
\end{figure}

We consider a tetrahedral chain in an external magnetic field, in which one Ising spin localized at the nodal lattice site regularly alternates with three equivalent lattice sites available for one mobile electron (see Fig.~\ref{fig1}). The Hamiltonian of the model looks as follows:
\begin{eqnarray}
{\cal H} &=& \sum_{k =
1}^N{\cal H}_k,\\
{\cal H}_k &=& -t\!\!\sum_{\alpha = \uparrow, \downarrow}\!\!\left(c_{k1,\alpha}^{\dag}c_{k2,\alpha} + c_{k2,\alpha}^{\dag}c_{k3,\alpha} + c_{k3,\alpha}^{\dag}c_{k1,\alpha} + {\rm h.c.}\right) \nonumber\\
&+& \frac{J}{2}\,(\sigma_{k}^{z} + \sigma_{k+1}^{z})\!\sum_{i = 1}^{3}\left(c_{ki,\uparrow}^{\dag}c_{ki,\uparrow} - c_{ki,\downarrow}^{\dag}c_{ki,\downarrow}\right) \nonumber\\
&-& \frac{H_{\rm I}}{2}\,(\sigma_{k}^{z} + \sigma_{k+1}^{z}) - \frac{H_{\rm e}}{2}\sum_{i = 1}^{3}\left(c_{ki,\uparrow}^{\dag}c_{ki,\uparrow} - c_{ki,\downarrow}^{\dag}c_{ki,\downarrow}\right).\nonumber
\label{eq:Hk}
\end{eqnarray}
\newline In above, $N$ is the total number of nodal lattice sites, $c_{ki,\alpha}^{\dag}$ and $c_{ki,\alpha}$ ($\alpha=\uparrow, \downarrow$) represent fermionic creation and annihilation operators, respectively, while $\sigma_{k}^{z}=\pm1/2$ stands for the Ising spins. The parameter $t>0$ takes into account the kinetic energy of a single mobile electron delocalized over a triangular plaquette and $J$ denotes the Ising exchange interaction between electrons and their nearest Ising neighbours. Finally, $H_{\rm I}$ and $H_{\rm e}$ are the Zeeman's terms acting on the localized Ising spins and mobile electrons, respectively.

The investigated spin-electron model is exactly solvable within the framework of a generalized decoration-iteration mapping transformation~\cite{4}. By the use of this rigorous procedure, one directly derives the mapping relation between the partition function of the investigated spin-electron tetrahedral chain and the partition function of the uniform spin-$1/2$ Ising linear chain in the effective external magnetic field, see Eq.~(2) in Ref.~\cite{3}. Subsequently, one can readily calculate the Gibbs free energy of the system and all important physical quantities utilizing the standard thermodynamic relations.

\section{Thermodynamics}
\label{results}
In this article, we turn our attention to the thermodynamic properties of the spin-electron tetrahedral chain. Before doing so, however, let us briefly mention the results obtained in Ref.~\cite{3}. By assuming equal magnetic fields acting on the Ising spins and electrons $H = H_{\rm I}= H_{\rm e}$, the model exhibits the ferromagnetic (FM) ground state with a full alignment of the Ising spins and mobile electrons into the $H$ direction, if $J<0$, or it passes from two-fold degenerate antiferromagnetic (AF) ground state with the antiparallel alignment between mobile electrons and their nearest Ising neighbours to the FM one at the field $H/J =1$, if $J>0$. In addition, an enhanced MCE can also be observed during the adiabatic demagnetization. More specifically, the model with $J<0$  exhibits an enhanced MCE in the limit $H/|J|\to 0$ when the entropy is kept very small, while the model with $J>0$ exhibits an enhanced MCE when the entropy is sufficiently close to the value $S/2N = \ln[(1+\sqrt{5})/2]$.

\begin{figure}[hb]
\begin{center}
\vspace{-1.0cm}
\includegraphics[angle = 0, width = 1.0\columnwidth]{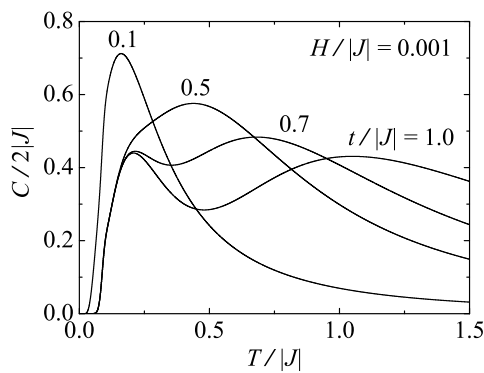}
\vspace{-1.0cm}
\caption{\small Thermal dependencies of the specific heat for the model with the ferromagnetic Ising interaction $J<0$ at the fixed magnetic field $H/|J|=0.001$ and various values of the hopping term $t/|J|$.}
\label{fig2}
\end{center}
\vspace{-0.25cm}
\end{figure}

Now, let us proceed to the examination of the thermodynamics of the spin-electron model with the ferromagnetic Ising interaction $J<0$. For this purpose, Fig.~\ref{fig2} illustrates thermal variations of the specific heat for this particular version of the model by choosing the magnetic field $H/|J|=0.001$ and different values of the hopping term $t/|J|$. As one can see, $C(T)$ curves exhibit at least a round high-temperature maximum, which diminishes and shifts towards higher temperatures upon strengthening $t$. For $t/|J|>0.5$, there also appears an additional low-temperature peak, which becomes more pronounced, the stronger the hopping term is. The observed double-peak structure of $C(T)$ curves originates from thermal excitations between the fully polarized ground state and the low-lying excited state with magnetic particles aligned in the opposite direction of $H$, which is energetically close to the ground state for weak fields. Indeed, the low-temperature maximum changes neither in the position nor height, if one considers the fixed $H$ and the parameter $t$ varies. However, when $H$ strengthens, then the low-temperature peak gradually shifts towards higher temperatures until it merges with the high-temperature one (it is not shown).

\begin{figure}[h]
\begin{center}
\vspace{-1.0cm}
\includegraphics[angle = 0, width = 1.0\columnwidth]{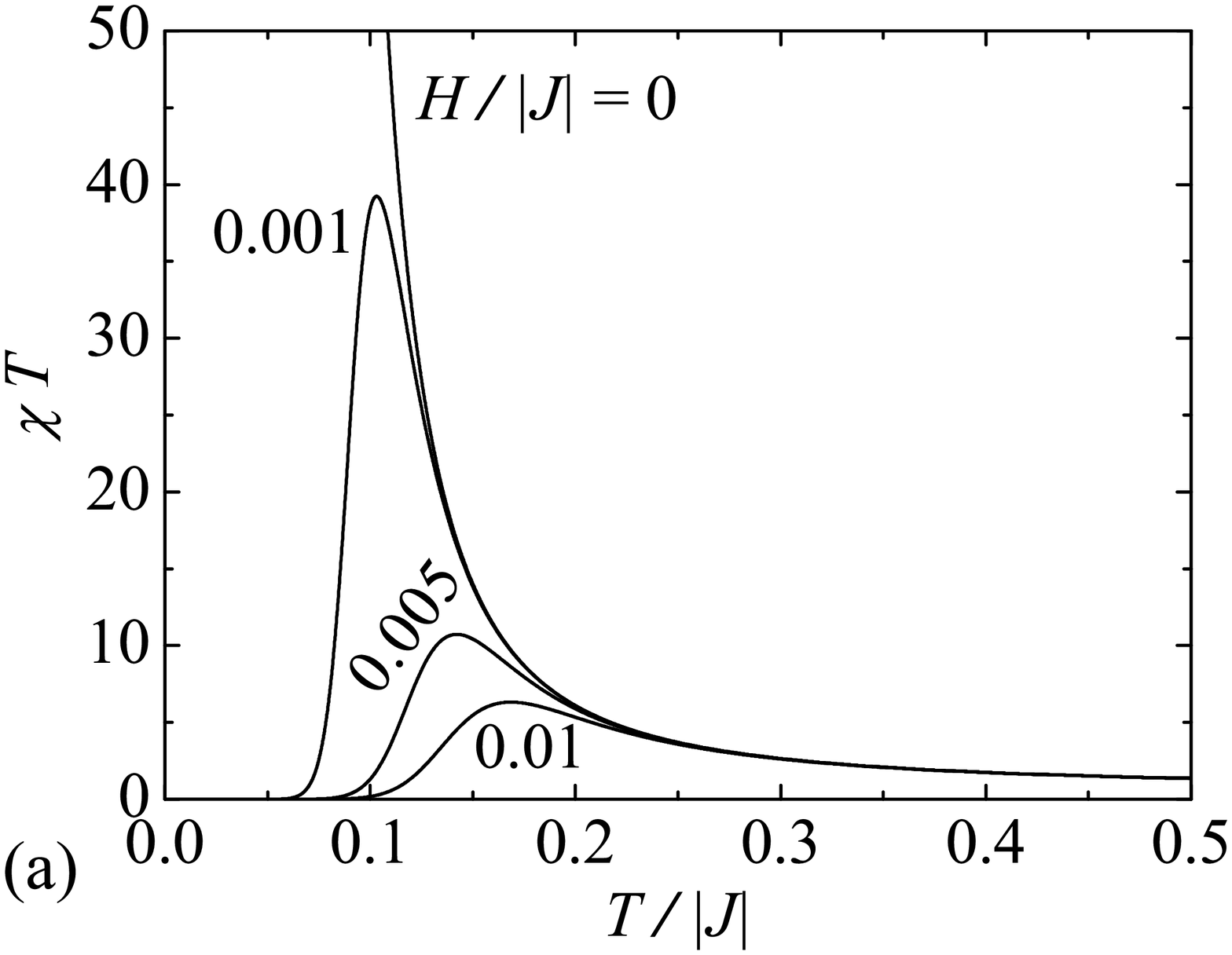}\\[-1.5cm]
\includegraphics[angle = 0, width = 1.0\columnwidth]{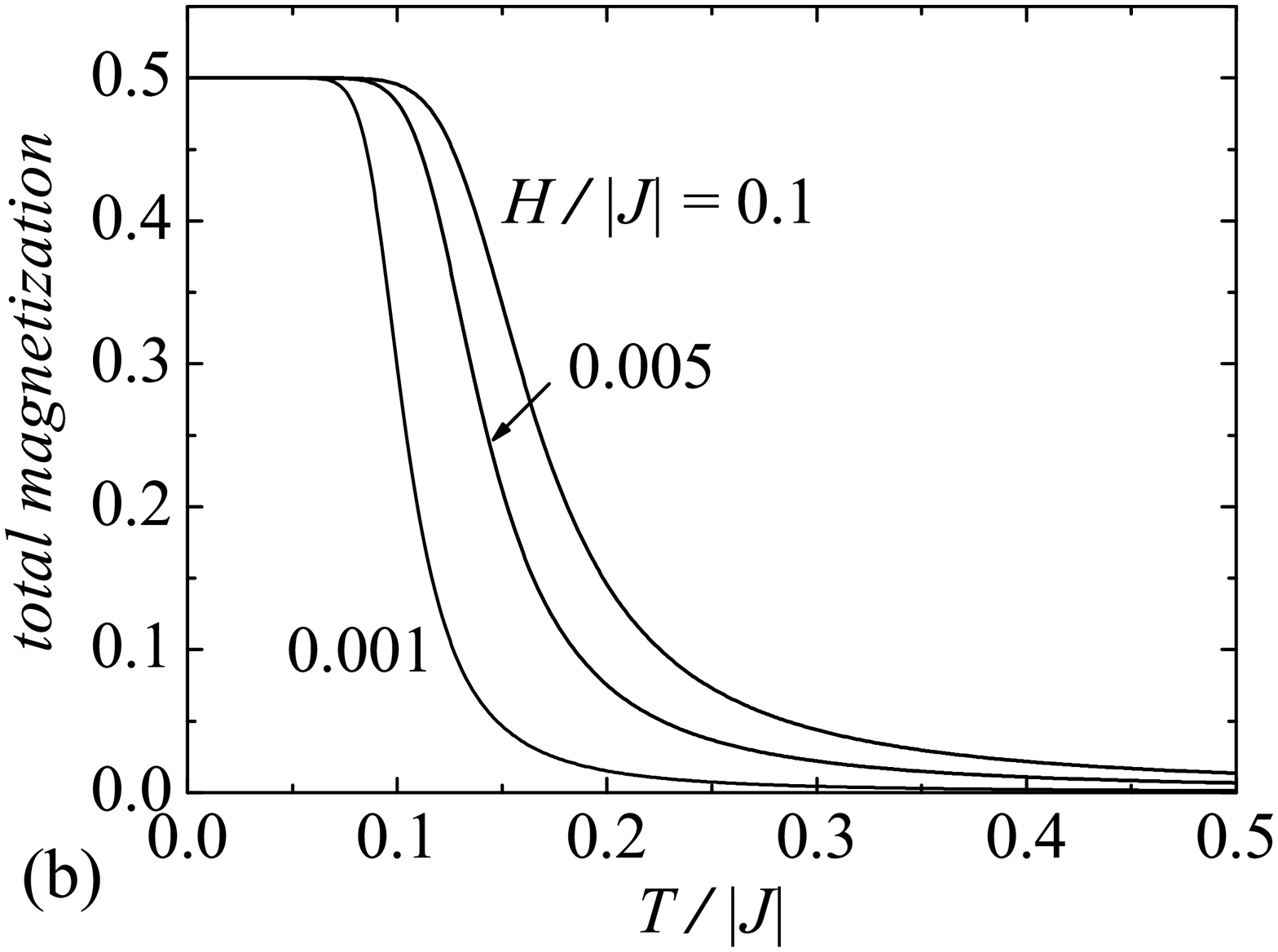}
\vspace{-1.0cm}
\caption{\small Thermal dependencies of the magnetic susceptibility times temperature [Fig.~\ref{fig3}(a)] and the total magnetization [Fig.~\ref{fig3}(b)] for the model with the ferromagnetic Ising interaction $J<0$ at various values of the magnetic field $H/|J|$.}
\label{fig3}
\end{center}
\vspace{-0.25cm}
\end{figure}

Another quantities, which are important for understanding the thermodynamics, are the magnetic susceptibility and the total magnetization. Thermal variations of the susceptibility times temperature ($\chi T$) are depicted in Fig.~\ref{fig3}(a) by assuming various values of the magnetic field. Note that the plotted $\chi T$ curves correspond to the particular case $J<0$ and any value of $t$. Obviously, the zero-field $\chi T$ product exhibits an exponential increase upon lowering temperature with low-temperature divergence that is typical for ferromagnets. By contrast, as far as the weak fields are considered, $\chi T$ plots start from zero value and manifest a maximum, which flattens and shifts to higher temperatures with the increasing $H$. The appearance of the peak in $\chi T$ curves relates to a thermal instability of the magnetically ordered system, whereas already a small temperature change necessitates a huge decrease of the total magnetization at weak fields, see Fig.~\ref{fig3}(b).

\begin{figure}[ht]
\begin{center}
\vspace{-1.05cm}
\includegraphics[angle = 0, width = 1.0\columnwidth]{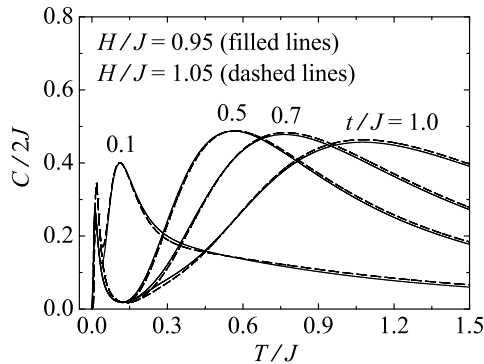}
\vspace{-1.0cm}
\caption{\small Thermal dependencies of the specific heat for the model with the antiferromagnetic Ising interaction $J>0$ at the fixed magnetic fields $H/J=0.95$ and $1.05$ by assuming various values of the hopping term $t/J$.}
\label{fig4}
\end{center}
\vspace{-0.25cm}
\end{figure}

\begin{figure}[h]
\begin{center}
\vspace{-0.5cm}
\includegraphics[angle = 0, width = 1.0\columnwidth]{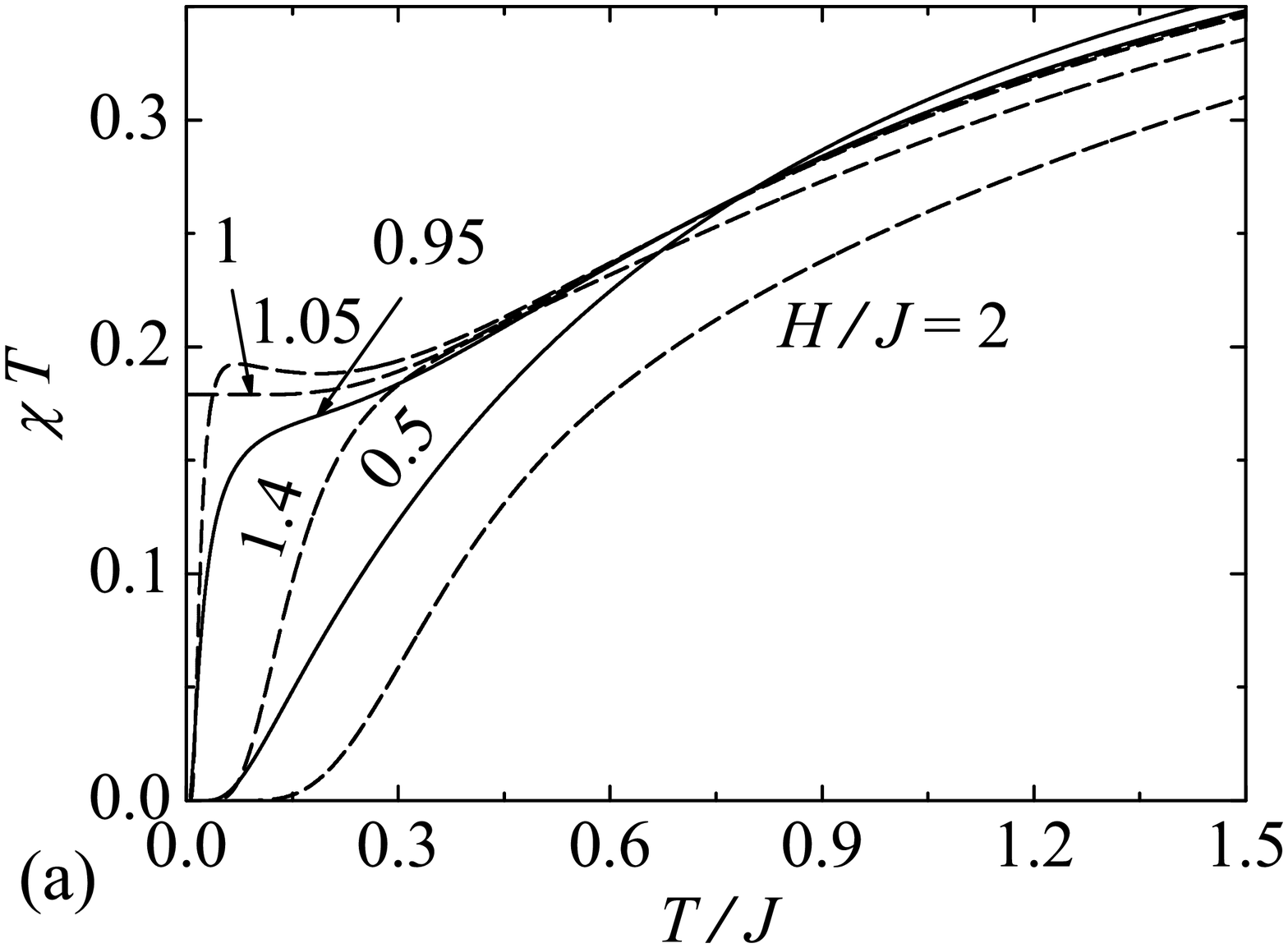}\\[-1.5cm]
\includegraphics[angle = 0, width = 1.0\columnwidth]{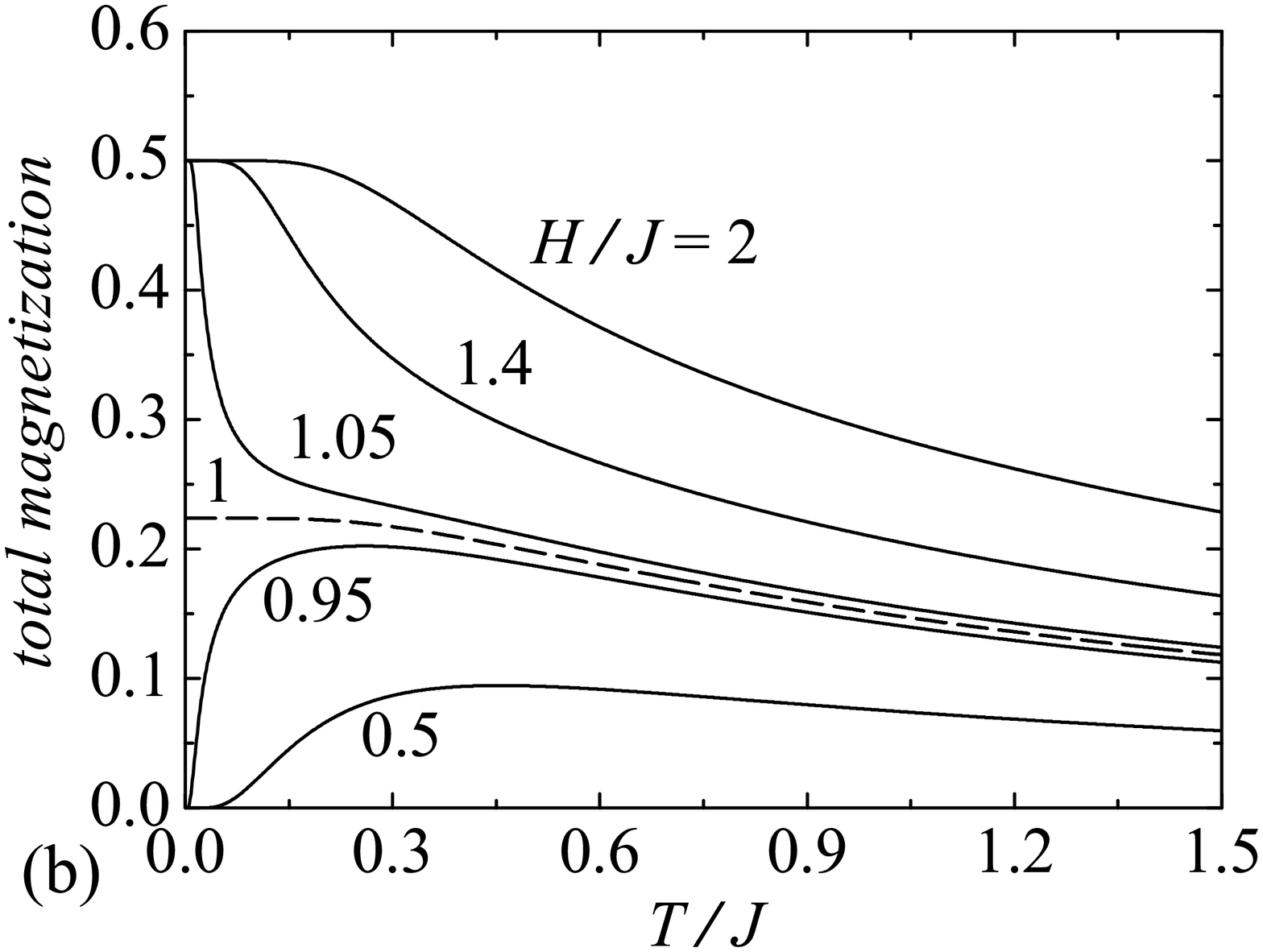}
\vspace{-1.0cm}
\caption{\small Thermal dependencies of the magnetic susceptibility times temperature [Fig.~\ref{fig5}(a)] and the total magnetization [Fig.~\ref{fig5}(b)] for the model with the antiferromagnetic Ising interaction $J>0$ at various values of the magnetic field $H/J$.}
\label{fig5}
\end{center}
\vspace{-0.5cm}
\end{figure}

To explore the thermodynamics of another particular case of the spin-electron tetrahedral chain with the antiferromagnetic Ising interaction $J>0$, the temperature dependencies of the specific heat, magnetic susceptibility and total magnetization are displayed in Figs.~\ref{fig4} and~\ref{fig5} for different values of the hopping parameter and magnetic field. Fig.~\ref{fig4} illustrates thermal variations of the specific heat when magnetic fields are chosen close to the transition value $H/J=1$, where the system undergoes a zero-temperature phase transition AF--FM. As one sees, $C(T)$ curves manifest a notable double-peak structure due to strong thermal excitations between ground state configuration and the one of the respective low-lying excited state, namely from AF towards FM configuration (for $H/J=0.95$) and vice versa (for $H/J=1.05$). The observed double-peak structure becomes more pronounced, the closer the magnetic field is selected to the value $H/J=1$ (it is not shown).

The aforementioned statements can be convincingly evidenced by thermal dependencies of the magnetic susceptibility multiplied by the temperature and total magnetization, which are displayed in Fig.~\ref{fig5} for various values of $H$. Similarly as for the model with $J<0$, plotted curves correspond to any value of $t$.  As can be seen form Fig.~\ref{fig5}(a), the quantity $\chi T$ exponentially tends to zero value with decreasing temperature regardless the magnetic field is chosen so that AF or FM phases constitute the ground state. The closer value of the external field to the AF--FM ground-state boundary is considered, the more pronounced temperature variation of the product $\chi T$ can be found due to the strengthening thermal excitations between energetically close AF and FM configurations. At the critical field $H/J=1$, there the magnetic susceptibility times temperature tends to the finite value $\chi T = 0.17889$ in the zero-temperature limit. In agreement with these observations, the total magnetization of the system exhibits the vigorous thermally-induced increase (decrease) from the zero value (saturated value) when the applied magnetic field is chosen slightly below (above) the critical value $H/J=1$, see Fig.~\ref{fig5}(b). The magnetization curve corresponding to $H/J=1$ slightly decreases from the non-saturated value $m=\sqrt{5}/10\approx0.22361$ with increasing temperature.

\section{Conclusion}
\label{summary}
In this paper, we have studied the thermodynamic properties of the exactly solvable spin-electron tetrahedral chain with Ising spins localized at nodal lattice sites and one mobile electron delocalized over the triangular plaquette. It has been demonstrated that the double-peak structure of the specific heat curves as well as pronounced temperature variations of the magnetic susceptibility multiplied by temperature and total magnetization observed around the fields $H/|J|=0$ (for $J<0$) and $H/J=1$ (for $J>0$) relate to strong thermal excitations between ground state configuration and the configuration corresponding to the respective low-lying excited state.
\\[4mm]
{\bf Acknowledgments}:
This work was financially supported by the grant of the Slovak Research and Development Agency under the contract No. APVV-0097-12.

\end{document}